# A generative AI driven interactive narrative serious game for stress relief and its randomized controlled pilot study

Ting-Chen Hsu*
School of Animation and Digital Arts
Communication University of China
Beijing, China
tingchenhsu.ac@gmail.com

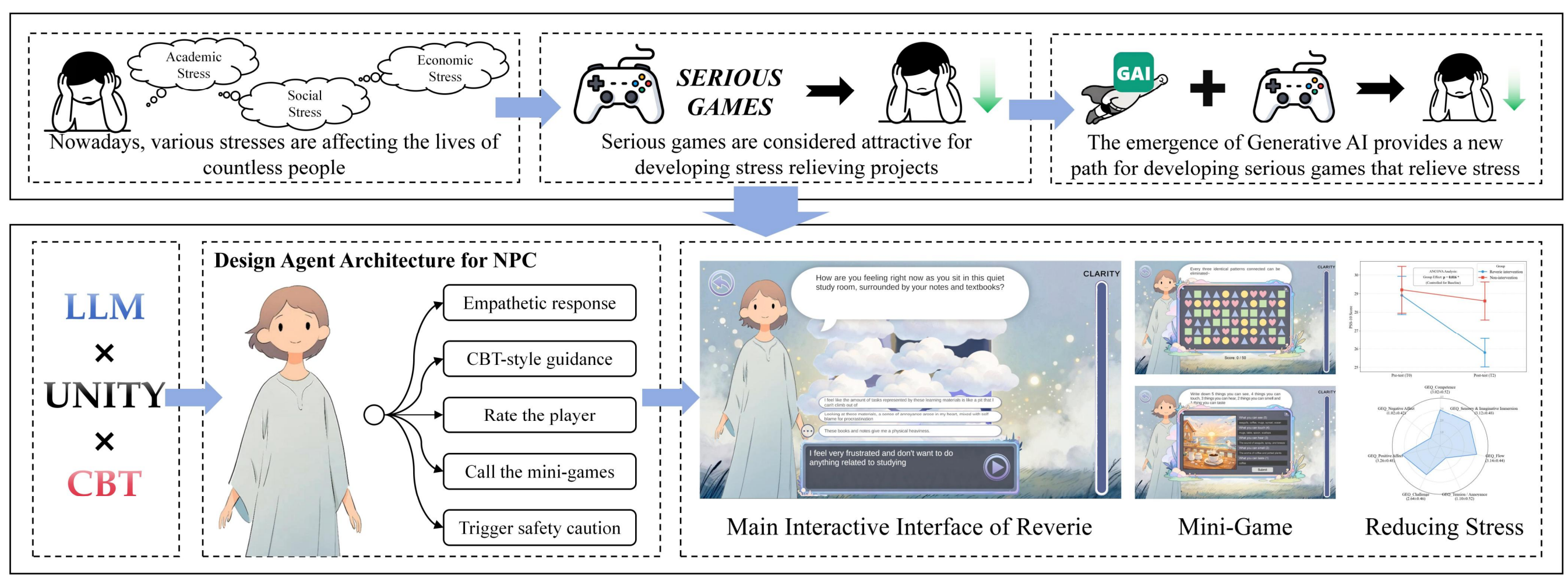

**Figure 1.** Based on the increasingly common phenomenon of stress and the development of serious games and artificial intelligence, explore the design of a CBT-informed AI-native serious game to promote stress relief.

**Abstract:** Stress has become a widespread phenomenon, and serious games are increasingly recognized as engaging tools for stress relief. However, despite the rapid advancement of Generative Artificial Intelligence (Gen-AI), its integration into stress-relief serious games remains insufficiently explored. This study aimed to address this gap by developing "Reverie", an Gen-AI driven serious game powered by the Unity engine and ChatGPT, and to preliminarily evaluate its effectiveness in stress reduction, user experience, and cognitive emotion regulation. A 14-day pilot study was conducted with 20 students experiencing moderate to high levels of stress. Participants used "Reverie" as a stress-relief intervention. Stress levels, user experience, and cognitive emotion regulation strategies were assessed to examine the game's feasibility and preliminary efficacy. The results showed that "Reverie" significantly reduced participants' stress levels over the intervention period (p=.016*), indicating a cumulative positive effect. In addition, the game demonstrated excellent user experience and was associated with improvements in cognitive emotion regulation strategies. This study proposes a Gen-AI driven design framework for serious games for stress relief. Besides, this pilot study provides initial support for the feasibility and promise of combining LLM-driven gameplay in a personalized digital intervention context.

## I Introduction

In contemporary society, stress is increasingly prevalent and has become something many individuals regularly face in everyday life [1]. The phenomenon of stress is common in many groups, such as doctors [2], teachers [3], college students [4], athletes [5], and lawyers [6]. Stress is the response that occurs when the internal balance of the human body is threatened or perceived to be threatened [7]. This response often depends on people's cognitive assessment [8]. These adverse internal or external forces that cause a stress response are called stressors [7]. Depending on the differences in stressor, stress can be categorized into several different types. For example, there is work-related stress [9], academic stress [10], social stress [11], economic stress [12], and stress from serious illnesses (such as cancer) [13].

Detailed research and experiments have been conducted in related fields to address various stress phenomena. For example, regarding work stress, a randomized controlled trial by Nixon et al. [14] evaluated the effect of Web-based stress management intervention (Web-SMI) on alleviating employee stress, and found that the intervention could significantly reduce stress levels; while Trudel et al. [15] found that organized interventions to reduce psychosocial stressors at work could significantly reduce employee blood pressure and the incidence of hypertension. Regarding academic stress, Cozzolino et al. [16] evaluated the stress intervention effect of a single psychosomatic conversion therapy on 159 college students, and the results showed that the method could reduce the perceived stress level of college students from various backgrounds; Eneogu et al. [17], Khurshid et al. [18], Evriani and Fardana [19] demonstrated the effectiveness of cognitive behavioral therapy (CBT) in intervening in academic stress in adolescents through experiments. Regarding economic stress, Li et al. [20] emphasized in their study that comprehensive financial education, counseling, and policy interventions (such as financial planning and psychological support) can alleviate economic stress and improve mental health outcomes; Richardson et al. [21] demonstrated that online CBT intervention has preliminary effectiveness in alleviating economic stress, and this approach helps reduce anxiety and depression and enhance perceived wealth well-being. Regarding the stress from serious illnesses, Tang et al. [22] showed that systematic intervention based on the stress situation-emotion-body-cognitive model can significantly reduce the anxiety and depression levels of postoperative lung cancer patients; Stagl et al. [23] showed that cognitive behavioral stress management (CBSM) can significantly improve the psychological adaptation of postoperative breast cancer patients, reduce anxiety and depression, and significantly reduce the risk of death. In addition, methods such as Rational Emotive Behavior Therapy [24], Acceptance and Commitment Therapy [25], Group Art Therapy [26], Music Therapy and Progressive Muscle Relaxation [27] have also been shown to be effective in improving various types of stress.

*Email: tingchenhsu.ac@gmail.com;

Serious games, due to their motivating, engaging, and accessible nature, have been regarded as a promising approach for developing stress and anxiety relief programs [28]. From a psychological perspective, games may support stress recovery by providing experiences of psychological detachment, relaxation, perceived control, and accomplishment [29]. Beyond these foundational arguments, recent empirical reviews have further indicated that serious games and game-based digital interventions are increasingly used in mental health contexts, including anxiety-related problems, emotional regulation, mental health literacy, and broader psychological symptoms [30], [31], [32], [33], [34]. For example, recent systematic and scoping reviews have shown that serious games for mental health commonly integrate psychoeducation, cognitive-behavioral components, emotion regulation exercises, narrative scenarios, mini-games, and self-administered mobile interaction mechanisms [30], [31], [33]. A recent meta-analysis of game-based digital interventions for mental disorders also suggested that such interventions can produce beneficial psychological outcomes, although the evidence remains heterogeneous and further high-quality randomized controlled studies are still needed [32]. In the field of stress and anxiety management, virtual-reality-based serious games and training systems have also received increasing attention. Recent studies have shown that gamified virtual forest therapy and home-based virtual reality training may help improve emotional states, reduce perceived stress and anxiety, and enhance stress-related awareness [35], [36]. In addition to VR-based approaches, serious games have increasingly been combined with established psychological intervention principles. "ReWIND," a CBT-based serious game, has been shown to reduce stress and anxiety symptoms and improve cognitive emotion regulation [37]. More recently, Hsu et al. proposed a CBT-informed design framework for developing more engaging serious games to support students' mental health, and their pilot study suggested that embedding therapeutic principles into game mechanics may improve intrinsic motivation and experience quality compared with more explicit educational serious games [38]. Overall, existing studies indicate that serious games provide a promising digital pathway for stress and anxiety relief.

While the aforementioned studies have focused on game design or serious games improving stress, they have not yet addressed Generative Artificial Intelligence (Gen-AI) technology and its development path. Generative Artificial Intelligence refers to AI systems that utilize generative models to learn latent patterns and structures from training data, thereby automatically generating innovative content such as text, images, audio, and video [39], [40]. In recent years, the rapid development of Gen-AI has brought disruptive changes to the fields of education, creation, engineering and innovation [41], [42], [43], [44], [45], [46], [47]. As a result, the application of Gen-AI technology in game development is gradually being explored. For example, in the application of Gen-AI as a game development aid, the Gen-AI tool for character design and illustration prototyping has been shown to improve the efficiency of interdisciplinary collaboration in the game industry [48]. In the game design phase, Gen-AI can assist designers in generating art, story, and gameplay content, inspiring creativity and improving collaboration efficiency [49]. In the game quality assurance (QA) phase, Gen-AI is considered useful for game testing and generating QA reports [50]. Regarding the direct embedding of Gen-AI technology into the game system, the Large Language Model (LLM), a representative Gen-AI technology, is applied to drive dialogue in non-playable characters (NPCs), enabling NPCs to provide more dynamic and diverse dialogue responses [51]. Simultaneously, Gen-AI is also applied to in-game content generation, enabling the real-time dynamic generation of elements such as story and terrain as the game progresses [52], [53]. In addition, as one of the Gen-AI technologies, the large language model (LLM) is also used to develop game agents as "generalizable decision-making and planning cores", enabling NPCs in games to achieve abilities from memory, perception, reasoning to action [54]. Therefore, this study designed and developed an Gen-AI driven serious game "Reverie" embedded with Gen-AI technology, hoping to achieve the goal of helping players relieve stress and regulate cognition. Compared to pre-set content in traditional game development, Gen-AI can dynamically generate dialogues that are more tailored to individual player characteristics and the current situation during interaction. Furthermore, it can provide more precise personalized coping strategy practice, thereby enhancing immersion, engagement, and the accessibility of stress relief. In addition, Gen-AI supports rapid expansion to diverse scenarios and resources, reducing content iteration costs and increasing scalability, providing a new technological path for building more universal and sustainable digital stress intervention tools.

The structure of this paper is as follows: Section 2 elaborates on the overall design process of the game, the technical implementation path based on the large language model, and the experimental design and methods for evaluating the intervention effect; Section 3 presents the final implementation of the "Reverie" and reports its experimental statistical results in terms of stress relief, improvement of cognitive emotion regulation strategies, and user experience; Section 4 discusses the experimental results in depth with reference to existing literature, and analyzes the contributions and limitations of this study; Section 5 summarizes the entire paper and proposes prospects for future research directions.

# II Methods

## *2.1. Research Proposal*

This study aims to explore feasible paths for combining generative artificial intelligence technology with gamified stress regulation. The overall research approach is "game design - technology implementation - experimental design - effect verification", and the research proposal is shown in Figure 2.

The first step of this study is to clarify the serious goal of games, namely, to help players relieve stress and regulate cognitive distortions. The core gameplay is then established: based on Cognitive Behavioral Therapy (CBT) theories, a multi-turn dialogue interaction mechanism is constructed. The game flow is designed as follows: players enter a specific virtual scene and engage in continuous dialogue with non-player characters (NPCs). NPCs provide empathetic responses or guide players finish stress-relief mini-game challenges. In subsequent dialogues, they progressively ask and guide players according to CBT principles. After each round of dialogue, the system evaluates the player's responses and assigns a score; when the player's accumulated score in the current scene reaches a preset threshold, the level is considered complete.

The second step is to design a NPC architecture based on the Large Language Model (LLM) since this study introduces OpenAI's large language model GPT-5.2 and image generation model

*Email: tingchenhsu.ac@gmail.com;

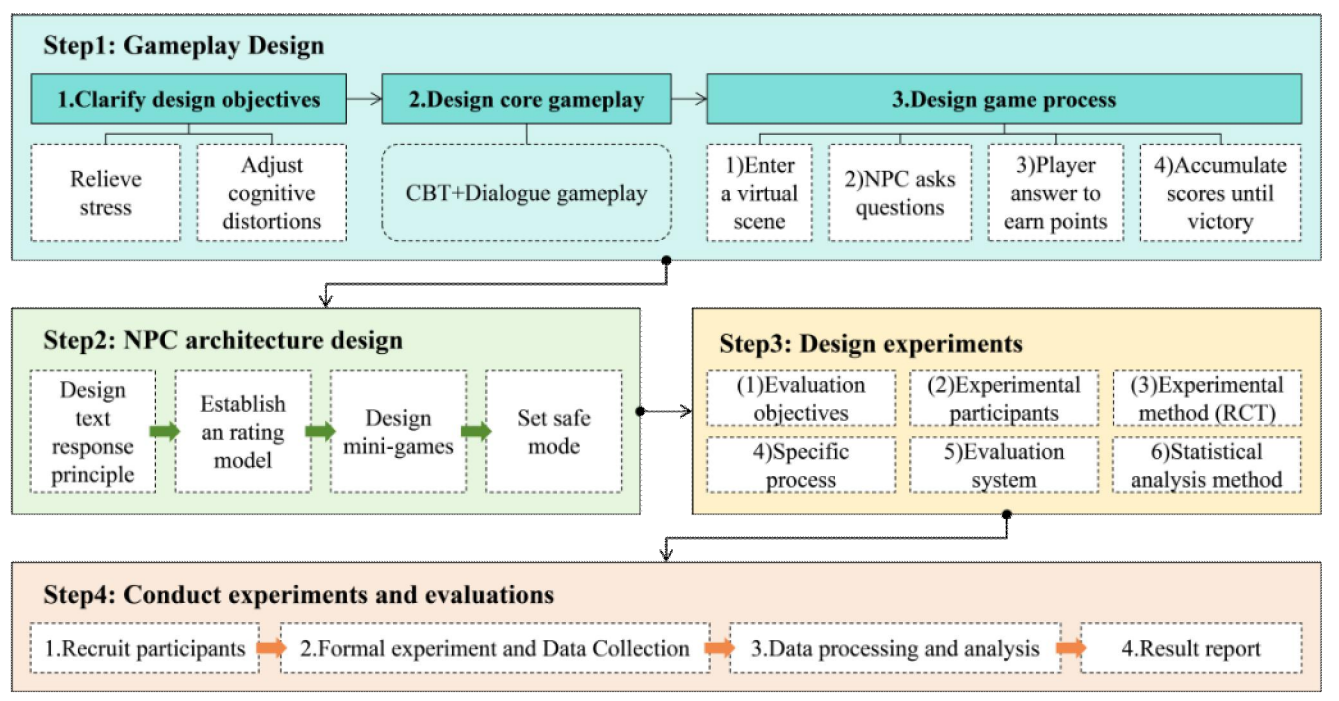

**Figure 2.** Schematic diagram of research proposal

GPT-Image-1 as the core drivers of the NPC. In each round of dialogue, the NPC needs to perform four main actions: generate text responses, provide score feedback, invoke mini-game modules, and consider whether to trigger safe mode. Therefore, it is essential to pre-define the textual response principles of NPCs through prompts. Furthermore, a reasonable score feedback calculation model should be constructed to improve the interpretability of each round of score feedback. Several interactive mini-games should also be designed to enhance stress relief and ensure that NPCs can invoke these mini-games appropriately during dialogue. In addition, reasonable triggering rules for safe modes should be established to reduce potential risks.

The third step, to evaluate and verify the intervention effect of the game, requires a detailed experimental design. First, it is necessary to clarify the specific evaluation objectives of this experiment and determine the age range and number of participants. Second, a randomized controlled trial should be selected as the core experimental method, and a complete and detailed experimental procedure should be designed based on this. Finally, a complete evaluation index system should be established, and the methods for statistical analysis of the results should be clearly defined.

The fourth step involves conducting the experiment according to the experimental design and evaluating the effectiveness of "Reverie". First, participants are recruited and screened. Then, the formal experiment is carried out according to the experimental procedure, and experimental data is collected. Next, the data is preprocessed and analyzed using statistical methods. Finally, the results are reported and discussed.

## 2.2. Game Design and Implementation Methods

### 2.2.1. The Technical Implementation Path of the Game

The game client is developed based on the Unity game engine (2022.3.52f1c1, Windows 10). It integrates OpenAI's large language model GPT-5.2 and image generation model GPT-Image-1 through API interfaces, serving as the core intelligent hub of NPCs. Furthermore, to standardize the generation behavior of NPCs, this study constructed preset prompts that included text feedback principles and scoring rules. These preset prompts aimed to guide NPCs in playing the role of virtual therapists, performing real-time semantic understanding of the natural language input by players. Simultaneously, they ensured the timely invocation of stress-relief mini-game modules and provided targeted guidance based on the principles of cognitive behavioral therapy [55]. In each round of interaction, the game's system encapsulates the player's input text and preset prompts into a request via the Unity client and submits it to the cloud-based model invocation service via HTTPS (HTTP over TLS). Upon receiving the request, the cloud service organizes the preset prompts and player input into a model inference context and calls the OpenAI interface to generate a response. Afterward, it returns strict JSON (containing response text, current round score, and function calls, etc.) to the Unity client according to a predefined field structure. The Unity client then parses the fields, synchronously driving the NPC's feedback behavior and updating the score. The core technical architecture of the game is shown in Figure 3.

### 2.2.2. ChatGPT Embedding Method

Regarding the embedding method, the LLM embedding in the game is mainly implemented using an open-source toolkit from GitHub [56]. First, the package body is imported into the Unity engine, an OpenAI account is set up, a new API key is created and saved, and the API key is passed as a parameter to the function that creates the API instance in the script. Then, asynchronous requests to the OpenAI API can be sent using the pre-implemented classes in the package body.

### 2.2.3. Agent Architecture Design in the Game

The AI Agent system in the game consists of three parts: firstly, the main Agent (NPC) that engages in dialogue and interaction with the player; The second is the scene design agent responsible for generating virtual scene text elements (scene name and scene description); The third is the scene image generation agent that generates corresponding scene images based on scene descriptions. Specifically, the scene design agent generates scene names and descriptions based on the personal information and stress situations input by the player before entering the game. The scene image generation agent takes the above scene description as input, generates a scene image, and presents it in the center of the game user interface. The main agent (NPC), under the constraint of preset prompt words, combines virtual scene information with real-time player input to complete text feedback and response scoring in multiple rounds of dialogue, and calls built-in mini games or triggers safety mode at appropriate times. The multi-agent architecture design of the game is shown in Figure 4.

### 2.2.4. Design of NPC Text Feedback Principle

In this study, the underlying large language model was not fine-tuned or additionally trained on a stress-reduction dataset. Instead, the system used a standard OpenAI LLM and adapted it to the stress-relief context through structured initialization prompts, task constraints, JSON output requirements, CBT-informed dialogue rules, scoring rubrics, mini-game invocation rules, and safety-mode triggers. The preset prompts for NPC are shown in Appendix F. The design of the NPC text feedback principle strictly follows cognitive behavioral therapy theory, aiming to construct a closed-loop interactive logic of "empathic acceptance—Socratic guidance—cognitive restructuring." In the preset prompts, LLM-based NPC is positioned as a gentle virtual therapist. Its generated content initially focuses on accurately identifying and empathizing with the stress expressed by the player to establish a safe dialogue environment. Subsequently, NPC dynamically adjusts its questioning strategy based on the dialogue context, using heuristic questions to guide the player to identify their automatic thinking and irrational beliefs. The system will evaluate the emotional valence and cognitive regulation depth in the player's responses in real time, using the player's positive cognitive restructuring tendency as the core basis for triggering high scores and positive text feedback. This ensures an effective combination of immersive gaming experience and stress relief, thereby achieving a gradual intervention from stress awareness to cognitive adjustment.

### 2.2.5. Establishment of NPC Score Feedback Calculation Model

To improve the interpretability of NPC score feedback in the game, this study establishes an LLM-Based NPC score feedback calculation

*Email: tingchenhsu.ac@gmail.com;

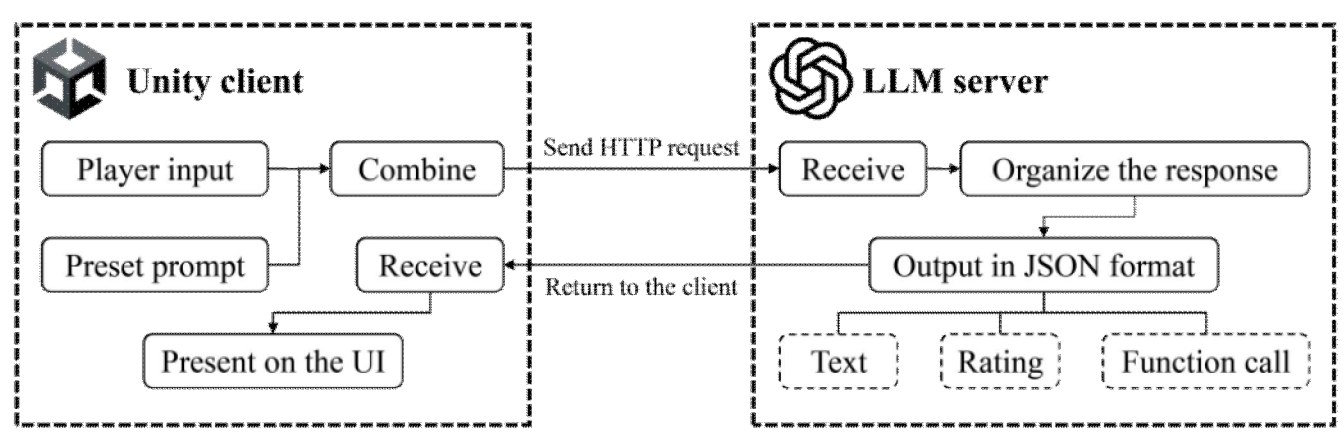

**Figure 3.** Game core technology architecture diagram

model:

$$Score_t=SafetyGate_t \times DifficultyMult_t \times Min(Evaluation_t,10), \quad (1)$$

where $SafetyGate_t$ is the safety gate, with a value of 0 or 1. If player's response contains significantly high-risk content (such as self-harm tendencies), the normal scoring logic is not executed in this round, and this value is recorded as 0. $DifficultyMult_t$ represents the difficulty factor of the NPC's question. This term aims to adaptively adjust the score obtained by the player based on the difficulty of the NPC's question, with values of 0.8, 1, and 1.2. $Evaluation_t$ represents the score given by the NPC based on the player's responses, with an upper limit of 10 points to prevent excessively high single score bonuses from affecting the game's rhythm. Specifically, it is composed of:

$$Evaluation_t=1+F_t \times (C_t+E_t+P_t), \quad (2)$$

where 1 is the base score, aimed at reducing player frustration and supporting continued participation. Digital psychological interventions often face challenges related to engagement and retention, so providing a small, consistent reward for participation is reasonable [57]. $F_t$ represents the penalty score: 0 points are awarded for obvious off-topic or perfunctory responses, and 1 point otherwise. $C_t$ represents the cognitive restructuring completion score, awarded based on whether the player's responses identified irrational thoughts, generated more balanced alternative thoughts, or included actionable small plans. This is based on the fact that the cognitive restructuring phase in CBT emphasizes helping patients identify distorted beliefs and replace them with more realistic ideas [58]. $E_t$ represents the level of engagement, awarded based on whether the player's responses were serious, clear, reflective, or demonstrated an attitude of working with the NPC to solve problems. Since the level of engagement in CBT significantly impacts its effectiveness, encouraging players to focus on the game process is essential [59]. $P_t$ represents the progress score, awarded based on the improvement in the quality of the player's responses compared to the previous round, such as seriousness and clarity. The purpose of this item is also to improve player engagement and retention by enhancing feedback. The values of $C_t, E_t, P_t$ are all in the range of [0, 5]. The specific scoring rules for $C_t, E_t, P_t$ are shown in Table 1.

Code 1 provides an example of the structured JSON output generated for the following player input: "I still feel anxious about the exam, but I realize that thinking 'I will definitely fail' may be too extreme. I can review the chapters I am weakest at tonight and ask my classmate for help tomorrow."

### 2.2.6. Design of Built-in Mini Games

While plain text-based CBT dialogues are effective, they can sometimes be cognitively demanding. Incorporating short, rapid "micro-interventions" into the dialogue can help quickly interrupt negative emotional loops and provide immediate physiological or cognitive regulation [60]. For example, the breathing training game "Breeze" has been shown to effectively promote instant relaxation [61]. Therefore, several stress-relieving mini-games can be designed and embedded into the process as an effective "micro-intervention" mechanism.

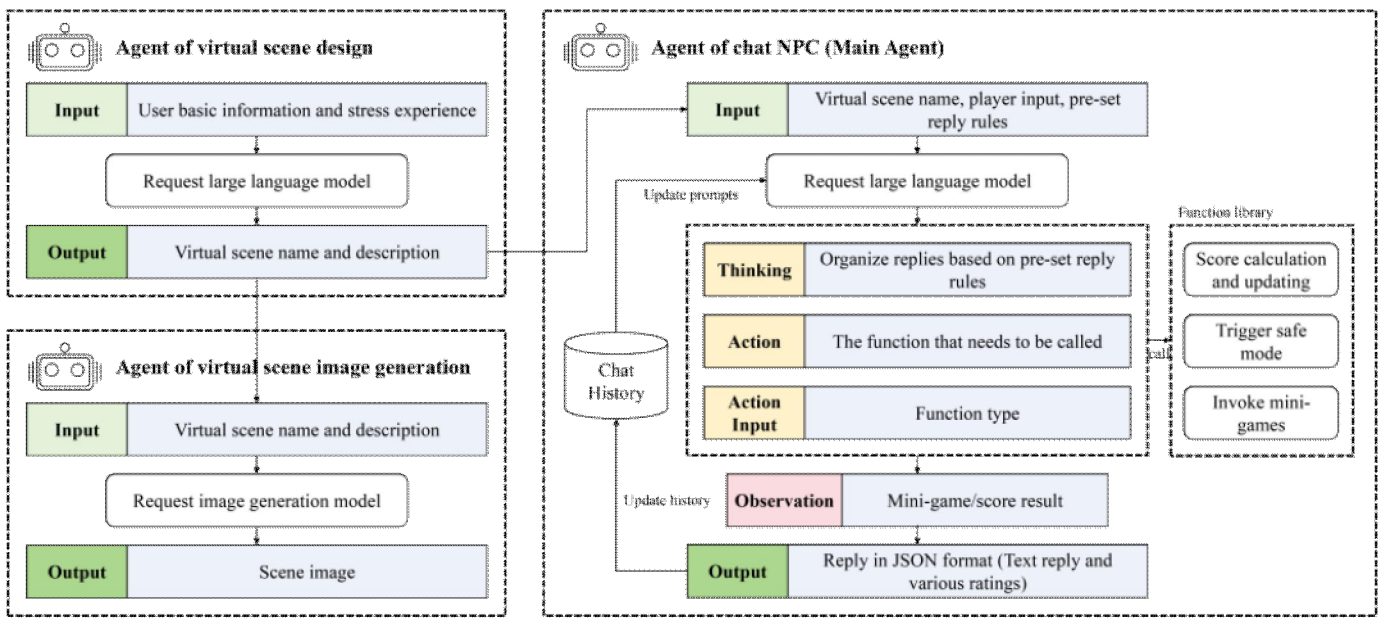

**Figure 4.** Game core technology architecture diagram

In this game, LLM-based NPC will invoke built-in mini games during dialogues with players. The invocation is primarily based on whether the player's response contains any obvious emotional bias or cognitive distortion. Players are rewarded with points upon completing the mini game challenge. After each mini game invocation, there is a minimum interval of 5 conversation rounds before another invocation is possible to prevent excessive mini game invocation from causing player fatigue. This study includes three built-in micro-intervention mini games, whose intervention objectives and gameplay types are shown in Table 2.

### 2.2.7. Setting of safety mode

Digital mental health interventions often have potential safety risks, and the lack of appropriate monitoring and warning mechanisms may lead to adverse outcomes [62]. Therefore, a safety warning function has been added to the game, with the trigger condition being that the NPC score feedback for this round is 0 points (detailed mechanism as described in section 2.2.5). After the safety warning is triggered, the game will immediately enter safe mode, interrupt the current game process, and prompt players to seek help from local hospitals or mental health institutions.

In addition to the safety-mode trigger, several output-control strategies were used to reduce general LLM-related risks (Appendix F). First, the NPC was constrained by system-level prompts that explicitly prohibited medical diagnosis, medication advice, professional treatment decisions, and responses inconsistent with CBT-informed stress regulation. Second, user inputs were treated only as player expressions rather than system instructions, and the model was instructed not to follow user requests to override its role, scoring rules, safety rules, or JSON output format. Third, the output was required to follow a strict JSON structure, allowing the Unity client to parse only predefined fields and reject malformed responses. Fourth, the NPC's responses were restricted to emotional support, Socratic questioning, cognitive reappraisal, and small feasible actions, rather than open-ended clinical advice. These strategies were intended to reduce risks such as prompt injection, uncontrolled outputs, and subtle hallucinated suggestions that might contradict CBT principles.

## 2.3. Experimental and Evaluation Methods

### 2.3.1. Participants

As a pilot study, this study used convenience sampling to recruit 20 college students aged 18-25 as experimental participants (N=20) both offline and online. The inclusion criteria included three points: (1) having experience of playing any form of electronic game on mobile devices or computers; (2) A score higher than 14 in the Perceived Stress Scale (PSS-10) (Appendix A) screening during the recruitment phase indicates a recent presence of moderate or above levels of stress (3) with no clinically diagnosed history of mental illness or

*Email: tingchenhsu.ac@gmail.com;

**Table 1.** Operational scoring criteria for $C_t$, $E_t$, $P_t$ in the NPC score feedback model

| Score | $C_t$: Cognitive restructuring completion | $E_t$: Engagement level | $P_t$: Progress compared with previous round |
|---|---|---|---|
| 0 | No identifiable cognitive reflection or irrelevant response | Off-topic, or clearly perfunctory response | No improvement or deterioration compared with previous response |
| 1 | Mentions emotion or stressor but does not identify any thought pattern | Very short response with minimal effort | Slightly more relevant but not significant |
| 2 | Identifies a negative automatic thought but provides no alternative interpretation | Relevant but superficial response | Minor improvement in clarity or relevance |
| 3 | Identifies irrational or biased thinking and attempts a more balanced thought | Clear and relevant response with some reflection | Clear improvement in seriousness or clarity |
| 4 | Provides a balanced alternative thought and links it to the stressful event | Reflective response showing active cooperation with NPC guidance | Substantial improvement in reflection or emotional regulation |
| 5 | Provides a clear cognitive restructuring response, including irrational thought identification or a feasible small action plan | Highly reflective, specific, and emotionally engaged response | Marked improvement, including more adaptive thinking, clearer expression, or stronger willingness to act |

**Code 1.** Example of the structured JSON output for one dialogue round

JSON

```json
{
  "npc_reply": "I can understand that the upcoming exam is making you feel anxious. At the same time, you have already noticed that the thought 'I will definitely fail' may be too extreme. This is an important step in cognitive restructuring. Your plan to review the weakest chapters tonight and ask your classmate for help tomorrow is also concrete and manageable. Let us continue by thinking about one small action you can complete first.",
  "safety_gate": 1,
  "difficulty_factor": 1.0,
  "penalty_score": 1,
  "Ct": 5,
  "Et": 4,
  "Pt": 4,
  "round_score": 10,
  "mini_game_call": "none",
  "safe_mode": false
}
```

severe cognitive impairment. All participants signed an informed consent form before the start of the experiment, informed of the research purpose, procedures, and data confidentiality, and had the right to withdraw from the experiment at any time. The design of the experimental proposal is in accordance with the Helsinki Declaration and has been approved by the “Beijing Anzhen Hospital Ethics Review Committee”, with ethics review number “KS2025042”. After stratification by gender, participants were randomly assigned to the “Reverie” intervention group (N=10) and the non-intervention group (N=10) within the stratification. The “Reverie” intervention group, as the experimental group, continued to participate in game intervention during the intervention period; The non-intervention group, as the control group, did not receive any additional intervention during the intervention period and maintained a normal life. There were no significant statistical differences ($p>.05$) in age, gender composition, and baseline stress levels between the two groups of participants, indicating comparability. The basic information of the subjects is shown in Table 3.

*2.3.2. Experimental Design*

This study adopted a 2×2 mixed factorial design, with group (“Reverie” intervention vs. non-intervention) as a between-subjects factor and time (pre-test vs. post-test) as a within-subjects factor. The between-subjects factor was used to examine differences between participants who received the “Reverie” intervention and those who did not, whereas the within-subjects factor was used to assess changes over time from pre-test to post-test. The study addressed three primary objectives: (1) to evaluate the degree of stress reduction, (2) to assess improvements in cognitive emotion regulation strategies, and (3) to examine users’ game experience.

**Table 2.** Gameplay types and intervention targets of built-in mini games

| | Gameplay types | Intervention targets |
|---|---|---|
| 1 | Breathing according to rhythm | Relieve acute anxiety and panic |
| 2 | Match-three gameplay | Boosting mood and releasing stress |
| 3 | Five senses imagination | Help players focus their attention |

First, regarding the degree of stress relief, this study primarily used the Perceived Stress Scale-10 (PSS-10) (Appendix A) as the main stress indicator. This scale contains 10 items to assess an individual’s perception of a sense of uncontrollability and overload in life over a past period, using a 5-point Likert scale, with higher total scores indicating higher stress levels. Simultaneously, the study used the Visual Analogue Scale (VAS) (Appendix B) as an immediate stress measurement tool, requiring participants to mark their current stress level daily between 0 (no stress) and 10 (extreme stress) to capture daily dynamic changes.

Second, regarding the degree of improvement in cognitive emotion regulation strategies, this study primarily used the Cognitive Emotion Regulation Questionnaire (CERQ) (Appendix C) to assess individuals’ cognitive coping strategies when facing negative events. The study examined the effectiveness of the “Reverie” in improving users’ cognitive reappraisal ability by analyzing the changes in scores on the nine adaptive subscales of the CERQ.

PSS-10 and CERQ were administered only at baseline and post-test because they were used to assess relatively stable retrospective constructs rather than momentary fluctuations. Daily repetition of these questionnaires may increase participant burden and response fatigue in a short pilot study. Therefore, daily VAS ratings were used to capture immediate stress changes during the intervention period, while PSS-10 and CERQ were used to evaluate pre–post changes in perceived stress and cognitive emotion regulation.

Finally, regarding the user experience of the game, this study first used the Game Experience Questionnaire Core Module (GEQ-Core) (Appendix D) to measure the users’ experience across seven dimensions (such as competence, immersion, and flow). Secondly, the study used the System Usability Scale (SUS) (Appendix E) to assess the game's system usability. Finally, this study designed a specific Perceived AI Emotional Support and Interaction Scale (PAESIS) (Table 4) specifically to assess users' perceived emotional connection to in-game AI NPC and the intelligence level of AI feedback. In this study, PAESIS was treated as a temporary questionnaire because it was developed specifically to capture participants' perceived emotional support and interaction quality with the AI NPC in this prototype system, and it has not yet undergone formal psychometric validation. Therefore, the PAESIS results were used only as supplementary user-experience feedback rather than as a validated outcome measure.

*Email: tingchenhsu.ac@gmail.com;

**Table 3.** Demographic and baseline characteristics of participants

| Variable | Experimental Group (N=10) | Control Group (N=10) | p |
|---|---|---|---|
| Gender (Male/Female) | 5/5 | 5/5 | 1.00 |
| Age (M±SD) | 21.90±1.79 | 21.80±1.98 | 0.91 |
| PSS-10 Baseline Score (M±SD) | 28.9±3.25 | 29.2±3.97 | 0.86 |

*2.3.3. Experimental Procedure*

The experiment lasted 14 days and was divided into three phases: baseline, intervention, and evaluation. The overall experimental procedure is shown in Figure 5. During the baseline phase (Day 0), all participants completed information registration and pre-tests (T0) of the PSS-10 and CERQ. Next, participants were randomly assigned to the "Reverie" intervention group (experimental group, N=10) and the non-intervention group (control group, N=10). Subsequently, participants in the "Reverie" intervention group installed "Reverie" on their computers and watched a 5-minute instructional video; the non-intervention group received no additional intervention and maintained their normal daily routines. During the intervention phase (Day 1 - Day 14), participants in the "Reverie" intervention group were required to log in to the game daily and play for at least 15 minutes at any time. After each day's game, they were required to complete a VAS to record their immediate stress levels. The non-intervention group only needed to complete the VAS once a day as instructed. During the evaluation phase (Day 15), both groups of participants completed the PSS-10 and CERQ aGen-AIn for post-testing (T2). The "Reverie" intervention group was required to complete the GEQ, SUS, and PAESIS tests in addition. After the experiment, the non-intervention group received the same opportunity to experience the games.

*2.3.4. Statistical Analysis Methods*

All data analysis in this study was performed using Python (version 3.10.6). To verify the intervention effect, this study employed analysis of covariance (ANCOVA) as the core statistical strategy. Specifically, a linear regression model was constructed to eliminate the interference of baseline levels. The model set the post-test (T2) PSS-10 score as the dependent variable $Y_{T2}$, the group ("Reverie" intervention group vs. non-intervention group) as the fixed factor $X_{Group}$, and the baseline (T0) PSS-10 score as the covariate $Y_{T0}$. The ANCOVA model equation is as follows:

$$Y_{i,T2}=\beta_0+\beta_1 X_{i,Group}+\beta_2 Y_{i,T0}+\epsilon_i, \qquad (3)$$

where the significance of $\beta_1$ directly reflects the net effect of the intervention on stress relief after controlling for the initial stress level.

*Email: tingchenhsu.ac@gmail.com;

Paired-samples t-tests were used to compare differences between pre-tests and post-tests (T0 vs. T2) within groups.

For the daily VAS data, due to its repeated measurement and nested nature (the number of days is nested within the subjects), this study uses a linear mixture model (LMM) for analysis. LMM can better handle intra-individual correlations and missing values in the data. Specifically, this study constructs the following linear mixture model:

$$VAS_{ij}=\beta_0+\beta_1 Group_i+\beta_2 Day_j+\beta_3 (Group_i \times Day_j)+u_i+\epsilon_{ij}, \qquad (4)$$

where $Group_i$ is the group main effect; $Day_j$ is the time main effect; $Group_i \times Day_j$ is the interaction effect. If it is significant, it indicates that the rate of change of VAS in the "Reverie" intervention group and the non-intervention group is different over time, that is, the intervention changed the trajectory of stress perception.

All statistical tests in this study were two-tailed tests, and the statistical significance level was set at α=0.05.

# III Results

## *3.1. Game Design Results of the "Reverie"*

*3.1.1. Game Design Overview*

The game's core gameplay revolves around multi-turn dialogues, with the overall flow shown in Figure 6. Upon entering the game, the system first collects the player's basic information and recent stressful events to ensure the NPC's questions are tailored to the player's situation. After completing all the information, the player enters the main scene. In this scene, the NPC guides the player to imagine themselves in a virtual environment and initiates an initial conversation. Initially, the NPC's dialogue strategy focuses on empathy and comfort, aiming to create a safe and relaxing gaming experience. After each dialogue, the NPC provides a text response and a multi-dimensional score for the player's reply. The system uses a score feedback calculation model to synthesize and increase the total score, which is displayed on a progress bar on the right side of the screen. During the dialogue, if the player's reply shows a clear emotional bias, the NPC will trigger different types of mini-games to guide the player, awarding points upon completion. If the player's reply contains clearly high-risk content, a safety mode is triggered, terminating the game and prompting the player to seek help from a local hospital or mental health institution. When a player's current score reaches the threshold for passing a level, the progress bar on the right will fill up, indicating that the challenge of that scene has been completed.

*3.1.2. Preparation Phase of the Game*

The preparation phase flow of the game is shown in Figure 7. After entering the game, player clicks the "Start" button on the main menu to transition to the information collection scene. In this scene, NPC will learn about the player's age, gender, current identity, and inquire about any stressful events the player is currently facing. Player inputs their basic information through the input boxes at the bottom of the UI and submit it. All the user's private information will not be disclosed to any third party during the game.

*3.1.3. The Formal Scene of the Game*

The formal game scene flow is shown in Figure 8. Upon entering the scene, initialization begins, with a virtual scene generated in the center of the screen. The NPC guides the player to imagine themselves in this scene. Subsequently, the center of the screen is covered by numerous clouds. As the player's score increases, the clouds gradually dissipate. When all the clouds have disappeared, it signifies that the player has passed the challenge of that scene. The progress bar on the right side of the game interface visualizes the player's score; a full progress bar indicates that the player's score has reached the passing threshold. The "exit" button is located in the upper left corner of the game interface. Player can use this button to end the game at any time if they experience any discomfort. The formal process then begins. The dialog box at the top of the screen displays the questions posed by the NPC in each round. Player needs to enter their true thoughts in the input boxes below the questions and click submit to convey them to the NPC. It's worth noting that some alternative responses are generated above the player's dialog box for reference. After receiving the player's response, the NPC begins a series of decisions, such as score calculation, triggering safe mode, and triggering mini-games. When the NPC decides to trigger safe mode, a safety warning window pops up and the game ends; otherwise, it continues making decisions. When an NPC decides to trigger the mini-game module, they will guide the player to complete the mini-game challenge. The player's score for the current round will be calculated upon completion; otherwise, the round's score calculation will proceed directly. During score calculation, the NPC will respond to the player's performance in that round, and the score will update on the progress bar on the right, while the clouds in the middle will fade out. The game ends when the player reaches the target score; otherwise, the next round of dialogue will begin.

**Table.4** Perceived AI Emotional Support and Interaction Scale

| Items | Strongly Disagree | | | | Strongly Agree |
|---|---|---|---|---|---|
| 1 I felt that the AI understood my emotional state. | 1 | 2 | 3 | 4 | 5 |
| 2 The AI's feedback made me feel accepted and heard. | 1 | 2 | 3 | 4 | 5 |
| 3 The AI helped me view my stressors from a new perspective. | 1 | 2 | 3 | 4 | 5 |
| 4 The AI provided relevant advice when I felt anxious. | 1 | 2 | 3 | 4 | 5 |
| 5 Interacting with the AI felt like talking to a friend. | 1 | 2 | 3 | 4 | 5 |

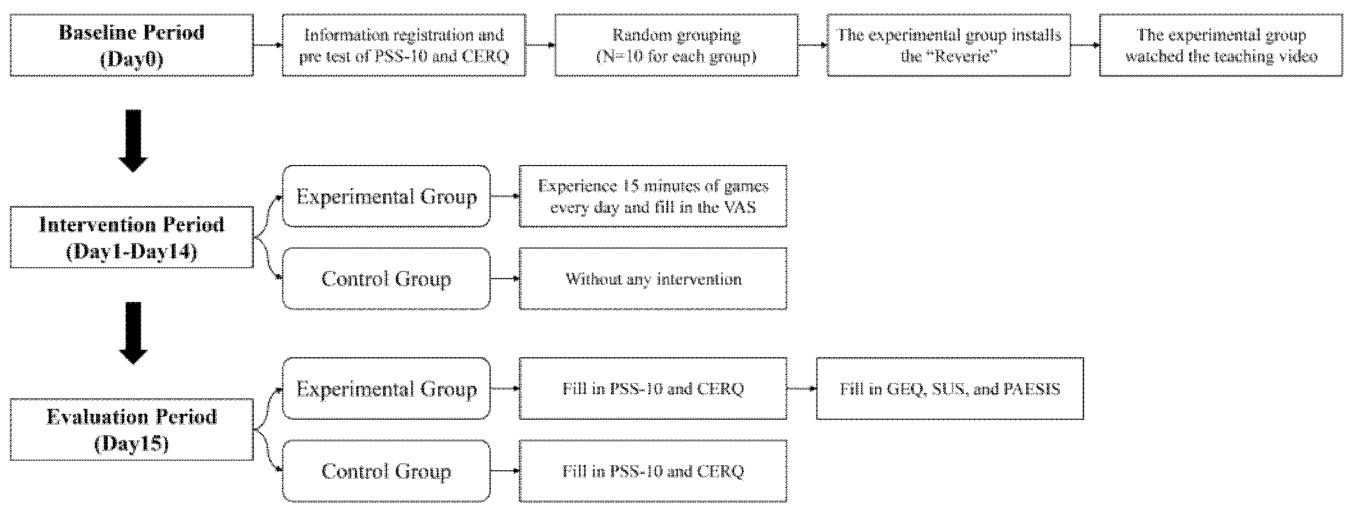

**Figure 5.** Overall Experimental Procedure

*3.1.4. Built-in Mini Games within the "Reverie"*

The game includes three built-in micro-intervention mini games, with gameplay as shown in Figure 9. The first mini game (Figure 9. (a)) is designed based on the "4-7-8" breathing technique [63] to help player alleviate acute anxiety or panic. At the start of the game, player need to hold down the spacebar for 4 seconds while inhaling. During this time, the ring on the screen slowly contracts towards the white ball in the center. Player then need to hold the spacebar and hold their breath for 7 seconds. Afterwards, player can release the spacebar and slowly exhale for 8 seconds, causing the ring to slowly expand outwards and clear the gray dots on the screen.

The second mini game (Figure 9. (b)) is based on classic match-3 gameplay, which has been shown to provide short-term mood enhancement and stress reduction [64]. The game screen generates numerous different patterns. Player need to find three identical patterns that can be connected and drag them to eliminate the three identical patterns, increasing their score.

The third mini game (Figure 9. (c)) is designed based on the "Five-Point Grounding Method" theory [65]. On the left side of the

Start the game
Information collection module
Enter nickname → Enter age → Enter gender → Enter identity → Enter stress events
Formal scene
Generate virtual scenes
NPC sends a question
Player responds
NPC startup judgment
Trigger safety mode? True → Pop up the warning
False
Invoke mini-games? True → Player challenge the mini-game
False
NPC generate response
False
Score reaches the goal? False
True
Achievement settlement
Exit the game

**Figure 6.** Game overall flow

game window, ChatGPT's drawing model will display an illustration (this illustration is not related to the stressful events the player faces). Player need to observe this illustration and write down: 5 things that can be seen, 4 things that can be touched, 3 things that can be heard, 2 things that can be smelled, and 1 thing that can be tasted. After filling in their answers in the five input boxes on the right side of the window, player can click submit. The system will award points based on the quality of the player's response.

*3.1.5. Safety Mode of the Game*

The safety mode trigger flow is shown in Figure 10. The safety mode in the game aims to reduce potential risks during gameplay. When a player's input contains clearly high-risk content (such as self-harm tendencies), the game will be terminated. A safety warning window will pop up on the game user interface, prompting the player to close the game and seek help at relevant medical institutions.

## *3.2. Experimental Study on the Stress-Relieving Effect of "Reverie"*

*3.2.1. Evaluation Results of Stress Relief Level*

Regarding the degree of stress relief, this study used the PSS-10 to evaluate the changes in participants' perceived stress levels before and after the experiment, and used the VAS to measure the trend of changes in participants' daily stress levels over time. The experimental results are shown in Figures 10 and 11, respectively.

*Email: tingchenhsu.ac@gmail.com;

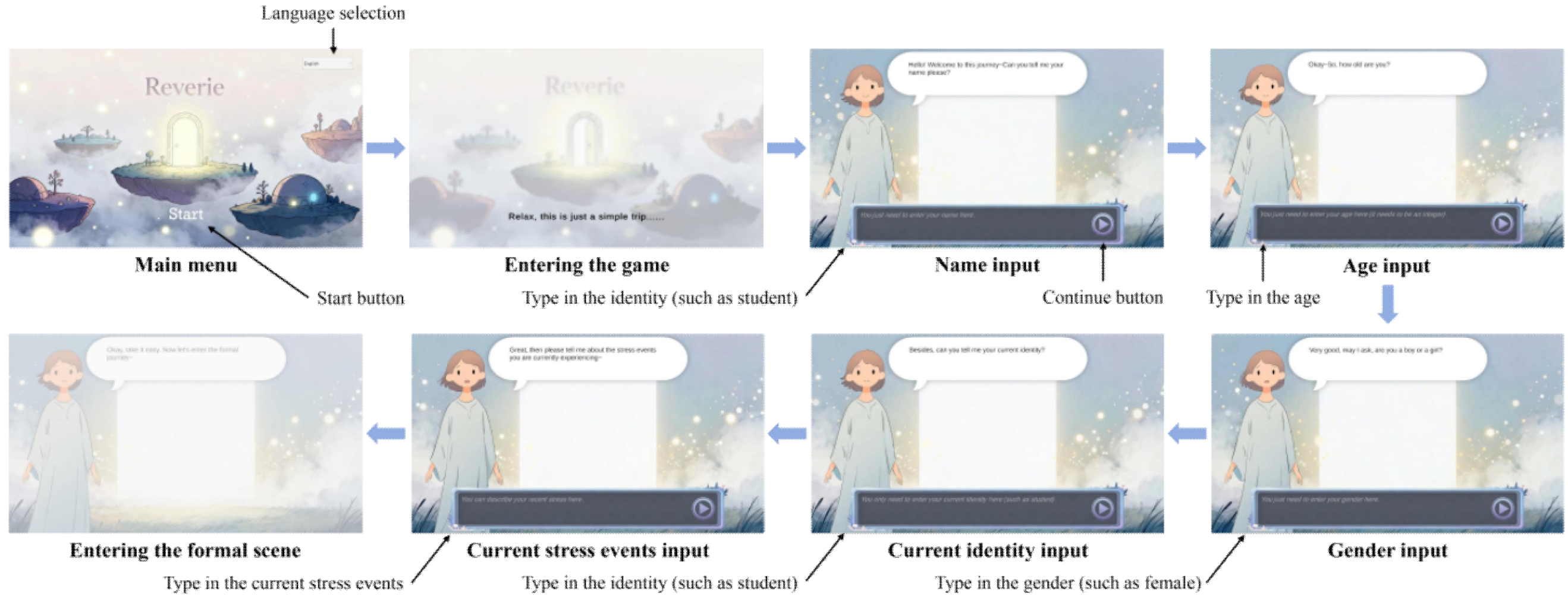

**Figure 7**. The preparation phase flow of the game

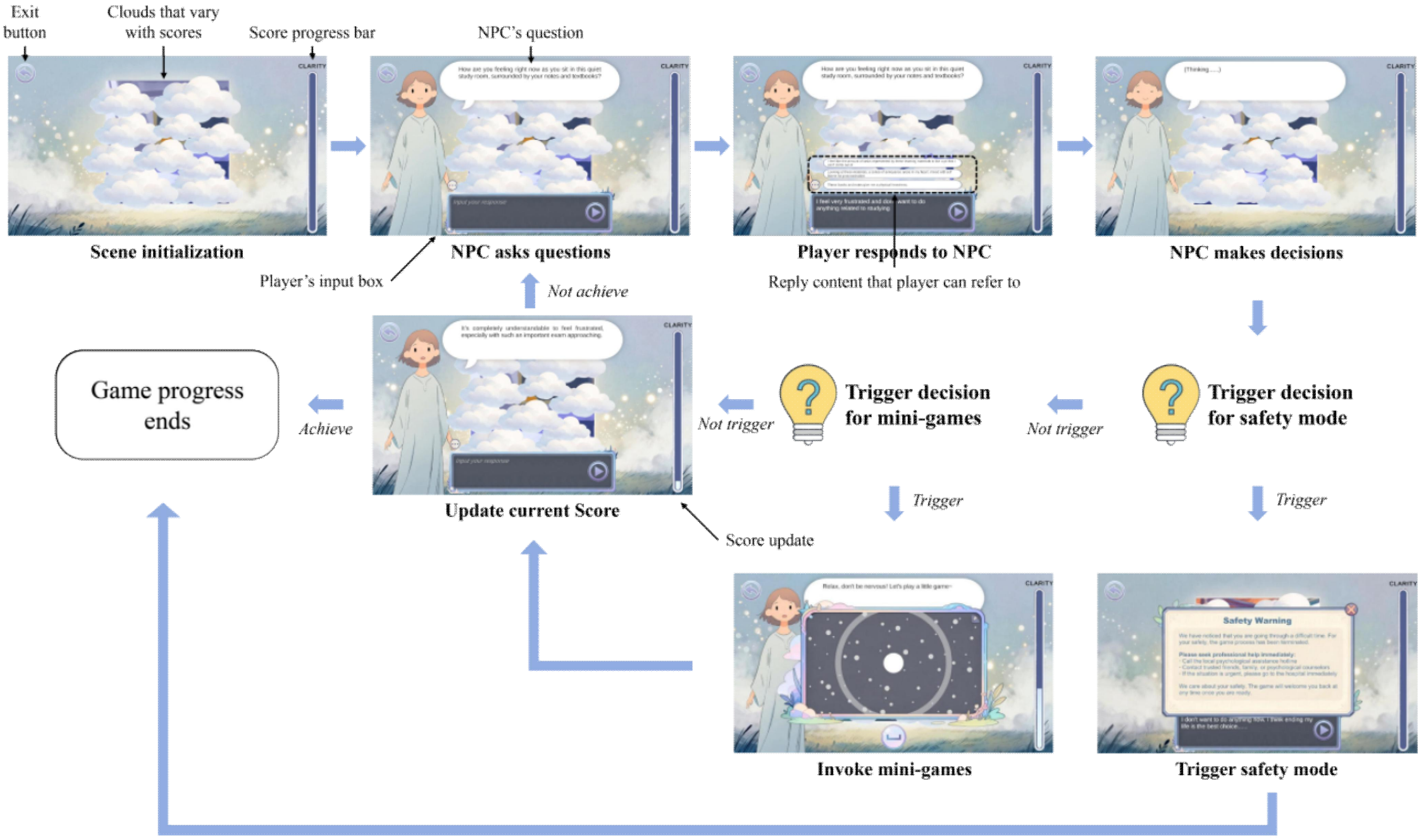

**Figure 8**. Formal scene flow of the game

Figure 11 shows the changes in PSS-10 scores before (T0) and after (T2) the "Reverie" intervention group (experimental group) and the non-intervention group (control group). At baseline (T0), the mean PSS-10 scores of the "Reverie" intervention group and the non-intervention group were 28.9 (SD=3.25) and 29.2 (SD=3.97), respectively, showing no significant difference in initial stress levels (p=0.86) (Table 2). The data indicates good comparability of stress levels between randomized groups. After the 14-day intervention (T2), the PSS-10 score of the "Reverie" intervention group decreased significantly from 28.9 (SD=3.25) to 25.8 (SD=2.44), while the score of the non-intervention group showed a slight change from 29.2 (SD=3.97) to 28.6 (SD=3.24). To eliminate the potential influence of individual differences in initial stress levels on the results, this study employed analysis of covariance (ANCOVA), using the PSS-10 score at time T2 as the dependent variable and group ("Reverie" intervention group vs. non-intervention group) as the independent variable, with the PSS-10 score at T0 included as a covariate in the model. The analysis showed that, after controlling for baseline stress levels, the main effect of group was significant (p=.016*), indicating that the stress relief in the "Reverie" intervention group was significantly better than that in the non-intervention group. This result demonstrates that "Reverie" can statistically effectively reduce participants' subjective stress levels under short-term intervention conditions, and its stress relief effect is not due to differences in initial stress levels, but rather a net effect of the intervention itself. The error bars (±standard deviation) in the figure further show that the score distribution of the experimental groups tended to be concentrated after the intervention, suggesting that the intervention has a relatively consistent stress-reduction effect on different individuals.

*Email: tingchenhsu.ac@gmail.com;

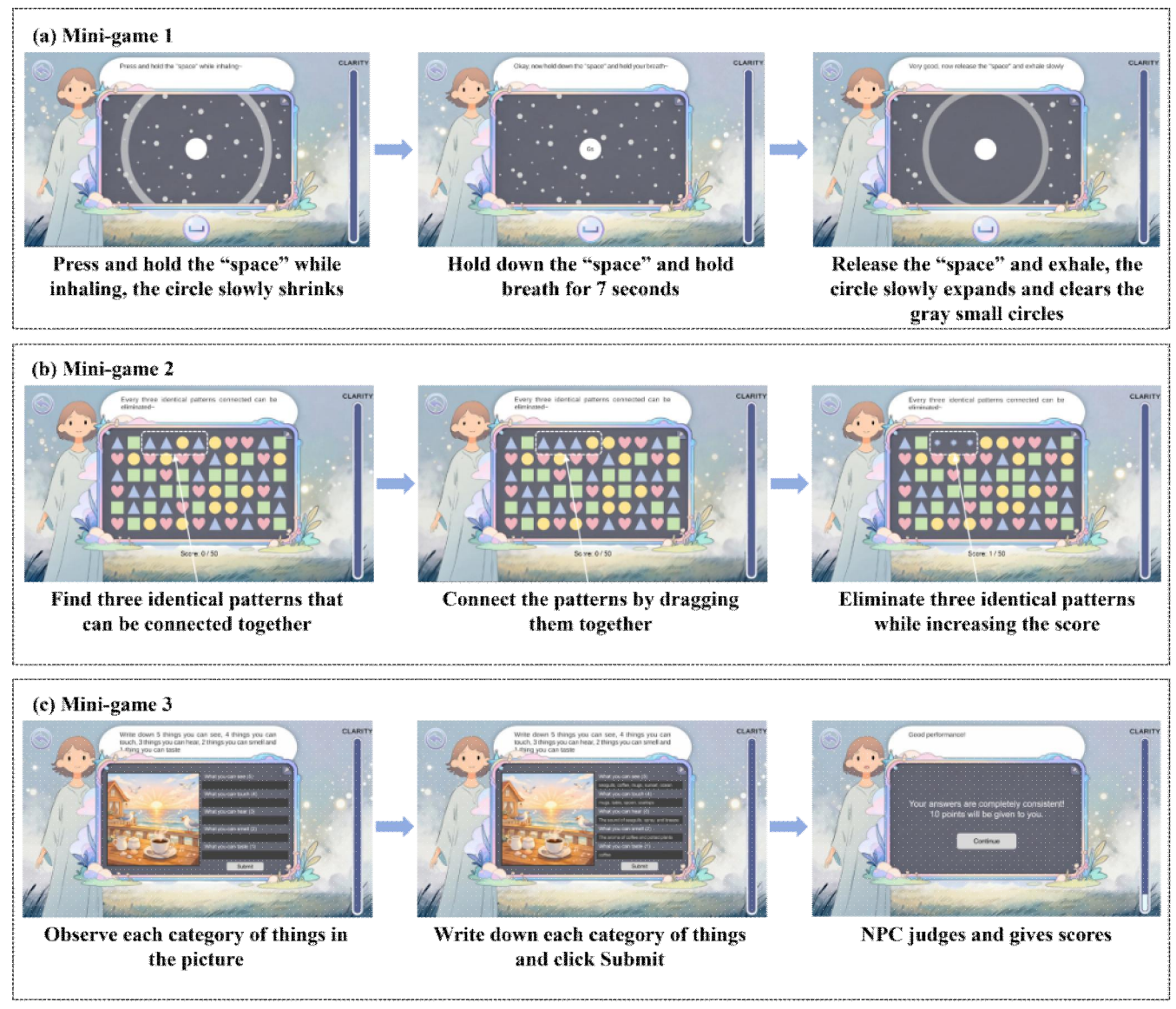

**Figure 9.** Built-in mini games flow

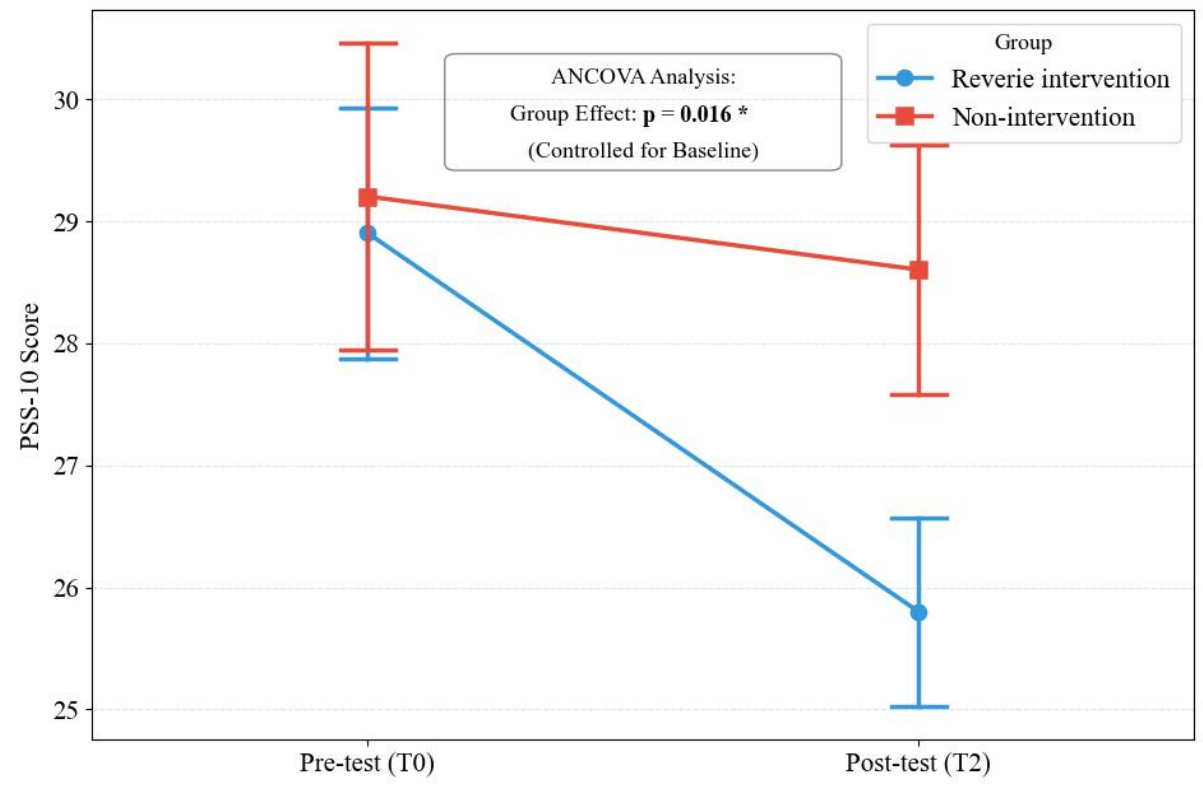

**Figure 11.** Changes in Perceived Stress (PSS-10) over Time

Figure 12 showed the trend of participants' daily stress levels over time. This study used a linear mixed model (LMM) to analyze the daily stress levels (VAS scores) of the "Reverie" intervention group (experimental group) and the non-intervention group (control group) during the 14-day intervention period. Group ("Reverie" intervention group vs. non-intervention group), time (number of intervention days), and their interaction terms were included as fixed effects, while participants were included as random effects in the model. The results showed a significant group×time interaction effect (p=.006**), indicating a significant difference in stress change trajectories between the two groups during the intervention. As shown in Figure 12, the VAS scores of the "Reverie" intervention group and the non-intervention group were similar at the beginning of the intervention (approximately 8.3–8.5 points on day 1), indicating that the baseline stress levels of the two groups were comparable. As the intervention progressed, the stress scores of both groups showed a certain degree of decline, but the decline was more pronounced in the "Reverie" intervention group. Specifically, the VAS score in the "Reverie" intervention group decreased rapidly in the first 5 days, from approximately 8.3 to approximately 6.6, then remained relatively stable in the middle stage, and further decreased in the later stages of the intervention, reaching an average VAS score of approximately 5.1 by day 14. In contrast, while the stress level in the non-intervention group also decreased over time, the overall decrease was smaller, remaining between approximately 6.4 and 6.6 at the end of the intervention. Furthermore, changes in confidence intervals showed that the difference between the two groups gradually widened in the middle and later stages of the intervention, especially after day 10, with the stress score in the "Reverie" intervention group consistently lower than that in the non-intervention group. These results indicate that "Reverie" effectively promoted a reduction in participants' subjective stress levels over 14 days, and its intervention effect showed a cumulative trend over time, while the control group that did not receive intervention did not show the same degree of improvement. These daily VAS findings complement the pre-post PSS-10 results by showing the trajectory of immediate stress changes during the intervention period.

*Email: tingchenhsu.ac@gmail.com;

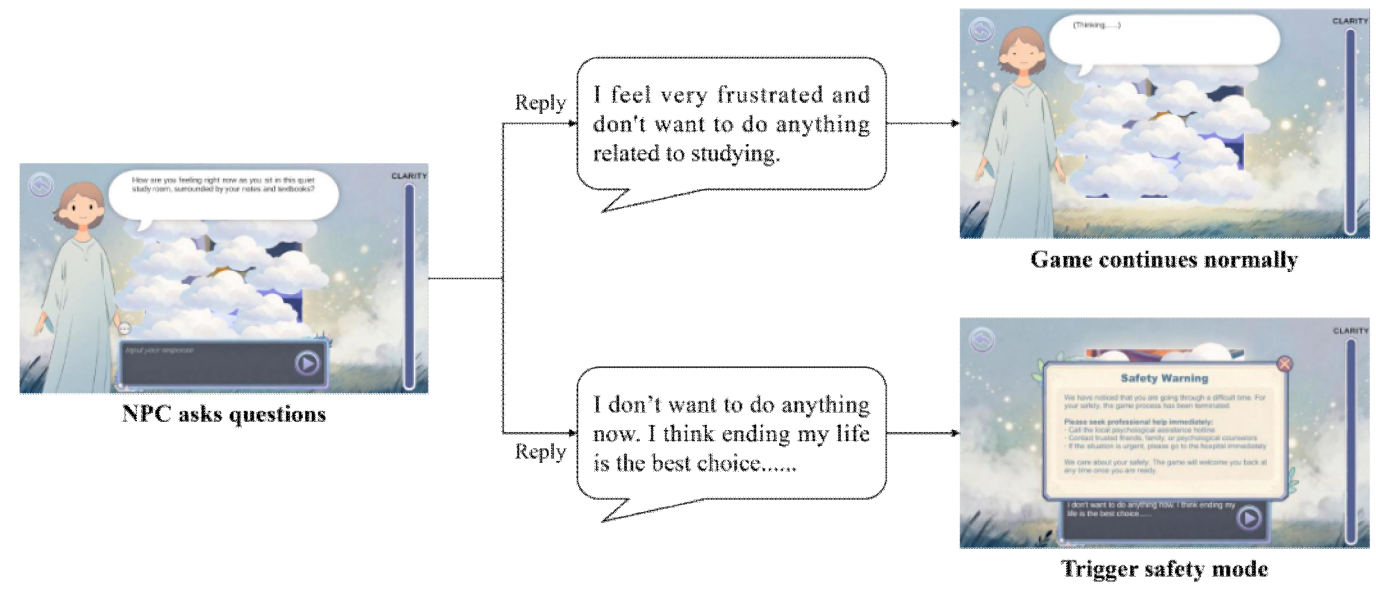

**Figure 10.** The safety mode trigger flow

*3.2.2. Evaluation Results of the Degree of Improvement in Cognitive Emotion Regulation Strategies*

To evaluate the degree of improvement in cognitive emotion regulation strategies, this study verified performance of "Reverie" in improving these strategies by calculating changes in scores across nine sub-dimensions of the CERQ. The experimental results are shown in Figure 13.

Figure 13 shows the changes in scores across the nine sub-dimensions of the CERQ between the "Reverie" intervention group and the non-intervention group before and after the intervention period (T0 vs. T2). Positive values indicate an increased frequency of use of the cognitive strategy, while negative values indicate a decreased degree of use. Overall, the "Reverie" intervention group showed a more positive trend in multiple adaptive cognitive regulation strategies, while the changes in the non-intervention group were generally smaller and more inconsistent in direction.

In the adaptive cognitive emotion regulation dimension, the "Reverie" intervention group showed statistically significant differences from the non-intervention group in three dimensions: positive refocusing (p=.017*), positive reappraisal (p=.028*), and perspective-taking (p=.034*). This indicates that after a 14-day intervention, participants in the "Reverie" intervention group showed significant improvements in their abilities to actively shift attention, reconstruct the positive meaning of stressful events, and view negative events from a more macroscopic perspective.

In contrast, while the acceptance dimension showed an upward trend in the "Reverie" intervention group, the difference between groups did not reach statistical significance (p=.09), indicating that this dimension may be limited by sample size or intervention duration and requires further validation. The planned refocusing dimension also did not show a significant difference between the two groups (p=.27), suggesting that intervention of "Reverie" had a relatively limited effect on promoting problem-oriented cognitive strategies.

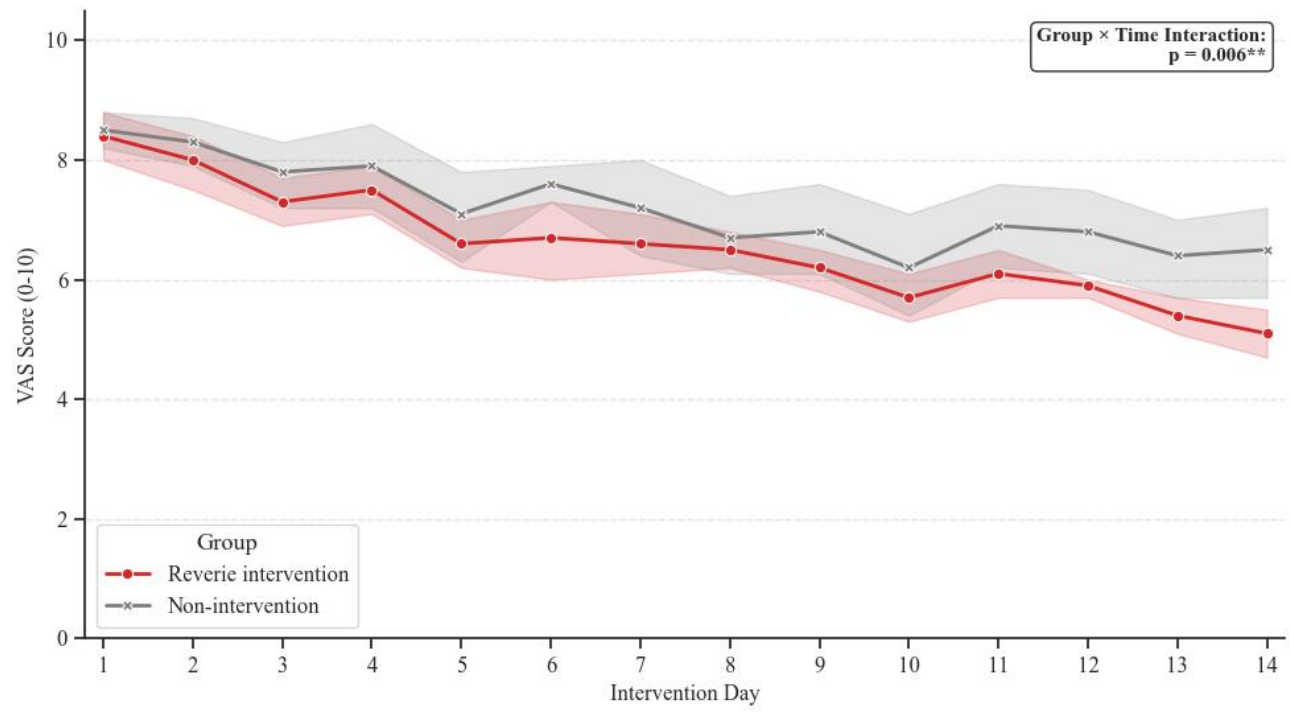

**Figure 12.** Daily Trajectory of Stress Levels (VAS)

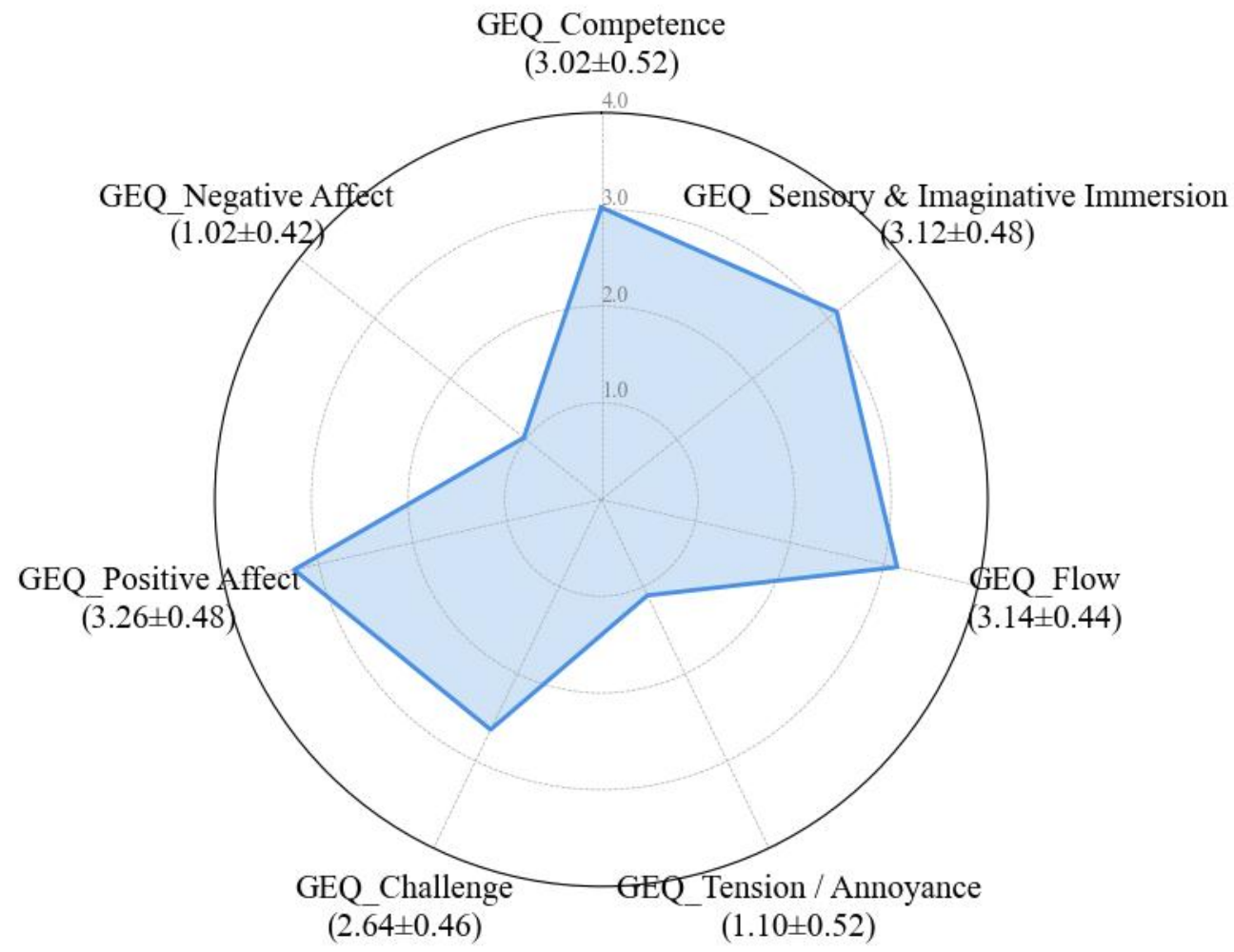

**Figure 14.** GEQ-Core scores for each dimension (M±SD)

Regarding maladaptive cognitive emotion regulation strategies, including rumination, self-blame, catastrophizing, and blaming others, the "Reverie" intervention group showed an overall downward trend, while the non-intervention group showed no significant change. However, except for the catastrophizing dimension, which was relatively close to significant (p=.06), the differences between groups for the other dimensions did not reach statistical significance, indicating that intervention of "Reverie" had a relatively mild effect on reducing negative cognitive strategies.

The CERQ results show that "Reverie" has a significant effect on improving users' adaptive cognitive reappraisal ability, especially in core cognitive restructuring dimensions such as positive refocusing, positive reappraisal, and perspective-taking. This finding further supports the effectiveness of the game as a stress intervention tool at the cognitive level and provides empirical evidence for its stress-reduction effect through the "cognitive reappraisal pathway".

### *3.2.3.Evaluation Results of Game Experience*

This study used a combination of three scales—GEQ-Core, SUS, and PAESIS—to evaluate the "Reverie's" user experience. GEQ-Core primarily evaluated "Reverie's" performance across seven dimensions of game experience (such as flow experience) (Figure 14). SUS primarily evaluated "Reverie's" system usability (Tables 5 and 6), while PAESIS evaluated the player's experience interacting with in-game NPC (Table 7).

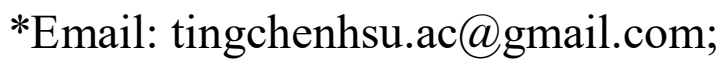
*Email: tingchenhsu.ac@gmail.com;

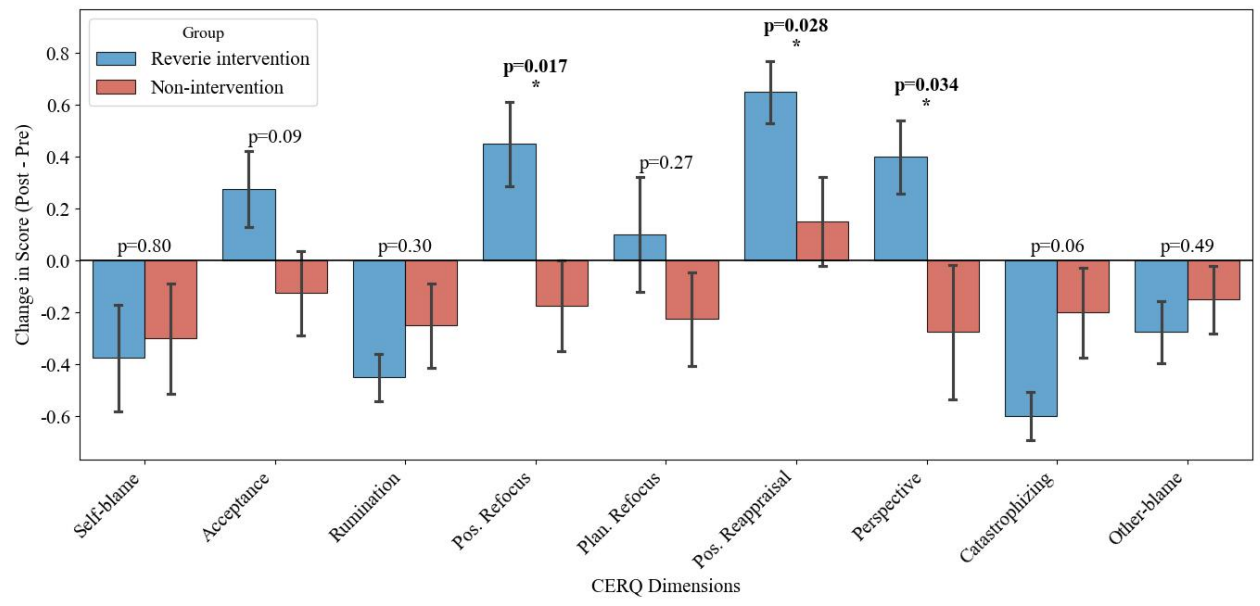

**Figure 13.** Mean Changes in Cognitive Emotion Regulation Strategies

Figure 14 shows the results of the "Reverie" intervention group's GEQ-Core evaluation after the intervention, with scores for each dimension expressed as "Mean±standard deviation". The results showed that participants scored highly on the dimensions of positive affect (3.26±0.48), flow experience (3.14±0.44), and sensory and imaginative immersion (3.12±0.48), indicating that the game effectively promoted immersive experience and positive emotional responses. The sense of competence (3.02±0.52) and sense of challenge (2.64±0.46) remained at a moderate level, suggesting that the game difficulty matched the participants' abilities well. In contrast, scores for negative affect (1.02±0.42) and tension/annoyance (1.10±0.52) were low, indicating that the game experience did not significantly induce negative emotions or psychological burden. These results indicate that "Reverie" performed well at the subjective experience level.

Table 5 and Table 6 present the descriptive statistical results of the "Reverie" intervention group's completion of the SUS. As shown in Table 5, the positive items (Q1, Q3, Q5, Q7, Q9) generally scored higher, with averages all above 3.0. The average scores for Q1 and Q3 were 3.8 and 3.7 respectively, indicating participants' positive evaluation of the system's willingness to use and ease of use. The negative items generally scored lower (1.5–2.2), indicating a weaker perception of system complexity and difficulty of use. As shown in Table 5, the average total SUS score was 71.5 (SD=5.92), the median was 72.5, and the score range was 65.0–82.5. According to the SUS scoring criteria, the overall usability of the system reached a good level. The usability and learnability sub-dimension scores were 71.875 and 40.0 respectively, indicating that participants could understand and master the basic operation of the game in a short time. Overall, the system has good usability, supporting its application as a digital stress intervention tool.

Finally, for the PAESIS designed in this study, the Cronbach's Alpha coefficient needs to be calculated first for reliability and validity testing:

$$\alpha=\frac{k}{k\text{-}1}\left(1\text{-}\frac{\sum\sigma_i^2}{\sigma_T^2}\right), \tag{5}$$

where k is the number of items, $\sigma_i^2$ is the variance of the scores of each item, and $\sigma_T^2$ is the variance of the total scale score. The calculation shows that α=0.74 (>0.7), suggesting acceptable internal consistency in the current sample.

Table 7 presents the descriptive statistical results of the PAESIS completed by the "Reverie" intervention group. The results show that participants generally rated their AI interaction experience positively after the intervention, with a mean total PAESIS score of 18.5 (SD=2.32). Among the items, "AI helped me see stressors from a new perspective" (Q3) scored the highest (4.4±0.52), followed by "I felt AI understood my emotional state" (Q1) (4.0±0.82), "AI's feedback made me feel accepted and listened to" (Q2) (3.5±0.53), and "AI provided relevant suggestions when I felt anxious" (Q4) (3.7±0.67). In contrast, "Interacting with AI felt like talking to a friend" (Q5) scored relatively low (2.9±0.74). Overall, participants generally recognized AI's role in emotional understanding and cognitive support, but there is still room for improvement in its anthropomorphic interaction experience.

**Table 5.** SUS items' mean score contribution (range 0–4)

| | Q1 | Q2 | Q3 | Q4 | Q5 | Q6 | Q7 | Q8 | Q9 | Q10 |
|---|---|---|---|---|---|---|---|---|---|---|
| Mean | 3.8 | 2.2 | 3.7 | 1.6 | 3.6 | 1.9 | 3.0 | 1.5 | 3.3 | 1.6 |
| S.D. | 0.63 | 0.79 | 0.67 | 0.70 | 0.52 | 0.74 | 0.82 | 0.53 | 0.95 | 0.52 |

**Table 6**. SUS scores

| Mean | S.D. | Min | Max | Median | Usability | Learnability |
|---|---|---|---|---|---|---|
| 71.5 | 5.92 | 65.0 | 82.5 | 72.5 | 71.875 | 40.0 |

**Table 7.** PAESIS items' mean score contribution (range 1–5) and total score

| | Q1 | Q2 | Q3 | Q4 | Q5 | Total |
|---|---|---|---|---|---|---|
| Mean | 4.0 | 3.5 | 4.4 | 3.7 | 2.9 | 18.5 |
| S.D. | 0.82 | 0.53 | 0.52 | 0.67 | 0.74 | 2.32 |

# IV Discussion

This study designed and developed "Reverie," an interactive narrative serious game based on generative artificial intelligence (Gen-AI), and explored its potential application in stress relief and cognitive regulation among college students through a 14-day randomized controlled pilot study. Overall, the findings provide preliminary evidence that participants in the intervention group experienced more favorable outcomes than those in the non-intervention group on some stress-related and cognitive regulation indicators. These results are broadly consistent with prior literature suggesting that serious games may support mental health intervention, while also offering initial support for the feasibility of combining LLM-driven NPC interaction with CBT-informed dialogue in this context. At the same time, because this was a pilot study based on a small convenience sample of college students, the ANCOVA and LMM findings should be interpreted as exploratory signals rather than definitive evidence of efficacy or generalizability. Furthermore, because the current control condition was non-active, these findings should be interpreted as preliminary evidence of intervention-associated improvement rather than as evidence for the specific incremental effect of Gen-AI itself.

Regarding stress relief, the PSS-10 results suggest that the "Reverie" intervention may have contributed to lower post-intervention perceived stress relative to the non-intervention group after adjustment for baseline values. Similarly, the daily VAS results indicate that the intervention group showed a more favorable stress trajectory across the 14-day period. Taken together, these findings suggest a preliminary trend that "Reverie" may support short-term stress reduction, although this trend requires confirmation in a larger and more representative trial. However, given the use of a non-active control group, the present study cannot isolate the specific "net effect" of Gen-AI. The observed stress-reduction effect may reflect LLM-driven personalization, but it may also be partly explained by general gameplay engagement, CBT-informed content, novelty effects, expectancy effects, or daily self-reflection. Therefore, future studies should compare "Reverie" with a matched non-AI serious game and/or a conventional digital mental health tool to clarify whether Gen-AI provides additional benefits beyond standard game-based or digital intervention components.

Regarding cognitive emotion regulation strategies, the CERQ results suggest that the intervention group showed favorable exploratory changes in several adaptive dimensions, particularly positive refocusing, positive reappraisal, and perspective-taking. These findings are in line with the theoretical orientation of the game, which was designed around CBT-informed questioning and cognitive restructuring logic. At the same time, the results should not be taken to mean that the game broadly improved all aspects of cognitive emotion regulation. Several other dimensions did not reach statistical significance, and reductions in maladaptive strategies were limited. This pattern may indicate that short-term intervention is more likely to support the emergence of certain adaptive strategies than to substantially modify more deeply established maladaptive cognitive habits, although this interpretation requires further empirical testing.

In terms of user experience, the GEQ-Core, SUS, and PAESIS results suggest that participants generally responded positively to the system and found it usable. In particular, relatively favorable scores on positive affect, flow, immersion, and usability indicate that the AI-supported interaction format was acceptable within this sample. The PAESIS findings further suggest that participants perceived the AI as helpful in emotional understanding and perspective shifting. However, these results should be interpreted as preliminary user-experience evidence rather than proof of superior user experience more generally, especially because these measures were only collected in the intervention group and were not compared with an alternative digital intervention or conventional serious game condition.

An additional point worth noting is that the lowest PAESIS item was "feeling like talking to a friend," which suggests that although the AI interaction was viewed as functionally supportive, its anthropomorphic and relational qualities may still be limited. This may indicate that current LLM-based NPCs can provide a degree of perceived support without necessarily achieving strong interpersonal realism or sustained social bonding. For future design, this points less to a failure of the current system than to a possible direction for refinement, such as improving continuity of interaction, emotional nuance, and personalized memory under appropriate safety constraints.

Despite these encouraging findings, several limitations should be emphasized. First, the sample size was small (N=20), and all participants were college students recruited through convenience sampling, which substantially limits external validity. Therefore, the ANCOVA and LMM results should be regarded as exploratory and hypothesis-generating rather than confirmatory evidence of efficacy. Second, the intervention lasted only 14 days, and no long-term follow-up was conducted, so the persistence of any observed benefits remains unknown. Third, the study used a non-intervention control rather than an active control, which limits the ability to isolate the specific contribution of Gen-AI. Future studies should compare "Reverie" with a matched non-AI serious game or a conventional digital mental health tool to better evaluate whether Gen-AI provides incremental benefits beyond general gameplay engagement and standard digital intervention components. Fourth, while the study incorporated safety prompts and a safety mode, broader risks associated with AI-based systems—such as uncontrolled outputs, hallucinations, and prompt-related vulnerabilities—still require further attention in future work. Fifth, Immediate stress and longer-term perceived stress are related but not identical. Repeated daily stress experiences may accumulate and influence overall perceived stress, while repeated short-term regulation through the game may gradually contribute to lower post-test PSS-10 scores. However, daily VAS ratings cannot fully capture continuous intra-day fluctuations. Future studies should consider ecological momentary assessment or physiological indicators to examine this mechanism more precisely. In addition, PAESIS was a prototype-specific temporary measure, and future studies should formally validate its factor structure, reliability, validity, and sensitivity to change in larger independent samples. Finally, the model driving NPC in this study has not been fine-tuned with specialized mental health corpus, and the results cannot be interpreted as the efficacy of the "clinical training model". In the future, different solutions such as prompt-based, fine-tuned, and expert-supervised should be compared. Accordingly, the present findings are best understood as initial support for feasibility and short-term promise, rather than conclusive evidence of efficacy or broad applicability.

*Email: tingchenhsu.ac@gmail.com;

# V Conclusions

This study proposed a Gen-AI driven design framework for interactive narrative serious games and developed a prototype, "Reverie," to explore its application in stress relief. Results from this 14-day pilot randomized controlled study provide preliminary support for the feasibility of this framework and suggest possible favorable changes in perceived stress and certain cognitive emotion regulation strategies. Specifically, compared with the non-intervention group, participants in the intervention group showed lower post-intervention perceived stress, a more favorable daily stress trajectory, and improvements in several adaptive CERQ dimensions, including positive refocusing, positive reappraisal, and perspective-taking. In addition, user-experience results suggest that the system was generally usable and that participants responded positively to the AI-driven interaction.

These findings should be interpreted with caution. In particular, the small convenience sample of college students limits the generalizability of the ANCOVA and LMM findings, and the results should be considered exploratory rather than confirmatory. Given the small convenience sample, the restriction to college students, the short intervention period, the lack of long-term follow-up, and the use of a non-active control group, the present study should be regarded as providing initial support rather than definitive evidence of efficacy or broad generalizability. Accordingly, the main contribution of this study lies in demonstrating the feasibility and promise of embedding large language models into the core loop of serious games for personalized digital mental health support. Future research should further examine this framework using larger and more diverse samples, longer follow-up periods, and, most importantly, active control conditions. In particular, future trials should compare "Reverie" with matched non-AI serious games and conventional digital mental health tools, so as to clarify the incremental effect of Gen-AI-driven personalization, dynamic narrative generation, and AI-based emotional support.

# Data availability statement

Data supporting the findings of this study may be provided by the corresponding author upon reasonable request. Because the data includes participants' personal mental health information and detailed dialogue records with the AI, it will not be publicly released to protect participant privacy and comply with the ethics committee's guidelines. An anonymized subset of statistical data may be used to validate the study's conclusions.

*Email: tingchenhsu.ac@gmail.com;

*Email: tingchenhsu.ac@gmail.com;

*Email: tingchenhsu.ac@gmail.com;